\documentclass[aps,twocolumn,showpacs,preprintnumbers,amsmath,amssymb,superscriptaddress]{revtex4-2}
\usepackage{graphicx}
\usepackage{dcolumn}
\usepackage{bm}
\usepackage{color}
\usepackage{amsmath}
\usepackage{amssymb}
\usepackage{latexsym}
\usepackage{multirow}
\usepackage{xcolor}
\usepackage{hyperref}
\usepackage[capitalise]{cleveref}
\usepackage{mathtools}
\usepackage[utf8]{inputenc}
\usepackage[T1]{fontenc}
\usepackage[super]{nth}
\usepackage{lmodern}
\usepackage{tabularx}
\usepackage{gensymb}
\begin{document}
\title{On the bridge hypothesis in the glass transition of freestanding polymer films}
\author{Haggai Bonneau}
\affiliation{Gulliver, CNRS UMR 7083, ESPCI Paris, Univ. PSL, 75005 Paris.}
\author{Maxence Arutkin}
\affiliation{School of Chemistry, Center for the Physics \& Chemistry of Living Systems, Ratner Institute for Single Molecule Chemistry, and the Sackler Center for Computational Molecular \& Materials Science, Tel Aviv University, 6997801, Israel.}
\author{Rainni Chen}
\affiliation{Department of Physics \& Astronomy, University of Waterloo, N2L 3G1, Ontario, Canada.}
\author{James A. Forrest}
\affiliation{Department of Physics \& Astronomy, University of Waterloo, N2L 3G1, Ontario, Canada.}
\author{Elie Rapha\"{e}l}
\affiliation{Gulliver, CNRS UMR 7083, ESPCI Paris, Univ. PSL, 75005 Paris.}
\author{Thomas Salez}\email{thomas.salez@cnrs.fr}
\affiliation{Univ. Bordeaux, CNRS, LOMA, UMR 5798, F-33400, Talence, France.}
\begin{abstract}
Freestanding thin polymer films with high molecular weights exhibit an anomalous decrease in the glass-transition temperature with film thickness. Specifically, in such materials, the measured glass-transition temperature evolves in an affine way with the film thickness, with a slope that weakly depends on the molecular weight. De Gennes proposed a sliding mechanism as the hypothetical dominant relaxation process in these systems, where stress kinks could propagate in a reptation-like fashion through so called bridges, \textit{i.e.} from one free interface to the other along the backbones of polymer macromolecules. Here, by considering the exact statistics of finite-sized random walks within a confined box, we investigate in details the bridge hypothesis. We show that the sliding mechanism cannot reproduce the basic features appearing in the experiments, and we exhibit the fundamental reasons behind such a fact.
\end{abstract}
\maketitle

\section{Introduction}\label{sec1}
Some liquids do not undergo a first-order phase transition to the crystalline solid state when being quenched in temperature, but rather exhibit a supercooled liquid-like behaviour with sharply increasing relaxation times as the temperature is reduced~\cite{Vogel1921,Fulcher1925,Tammann1926,Williams1955,Angell1995} -- a phenomenon known as the glass transition~\cite{Berthier2011,Ediger2012}. While recent theoretical breakthroughs have shown the existence of an ideal glass transition in infinite space dimensions~\cite{parisi2020theory}, a complete understanding of the formation of real glassy materials remains a central unsolved problem in condensed matter physics~\cite{Anderson1995}. In addition to the interest in this fundamental problem, glassy materials find widespread use and their rheology and stability have significant technological importance. 
	
In the supercooled liquid state, particles are crowded and must move in a correlated way to allow for a reorganisation or a relaxation event~\cite{Adam1965}. This phenomenology has been used to suggest the emergence of a dynamical cooperative length scale~\cite{Donth1996}. To probe this hypothetical length scale, besides bulk numerical simulations~\cite{Stevenson2006}, or mimetic jammed colloidal systems~\cite{Liu2010,Zhang2011}, an alternative strategy consisted in studying finite-size effects on the glass-transition temperature $T_\text{g}$, \textit{i.e.} through systems with a similar nanometric size as the cooperative length at stake~\cite{Bares1975,Jackson1990,Scheidler2000,Berthier2003}. In particular, experiments and numerical studies investigating thin glassy polymer films have been performed~\cite{Keddie1994, Forrest1996,Varnik2002, Ellison2003, Baschnagel2005, Alcoutlabi2005,Fakhraai2008,Yang2010,Chai2014,Yoon2014,Ediger2014}, and have revealed a set of rich and exotic phenomena. Most notably, a reduction of $T_\text{g}$ in thin films of many materials was observed and was further attributed to a combination of the dynamical correlation length with an enhanced liquid-like surface mobility in glasses. These observations have been studied from a theoretical point of view as well, but there is no definitive consensus yet on the exact underlying mechanisms at play~\cite{Ngai1998, Long2001, Herminghaus2001, Lipson2009, Forrest2013, Lam2013, Forrest2014, Mirigian2014, Salez2015, Hanakata2015, Arutkin2020}. 
\begin{figure} 
\includegraphics[width=0.45\textwidth]{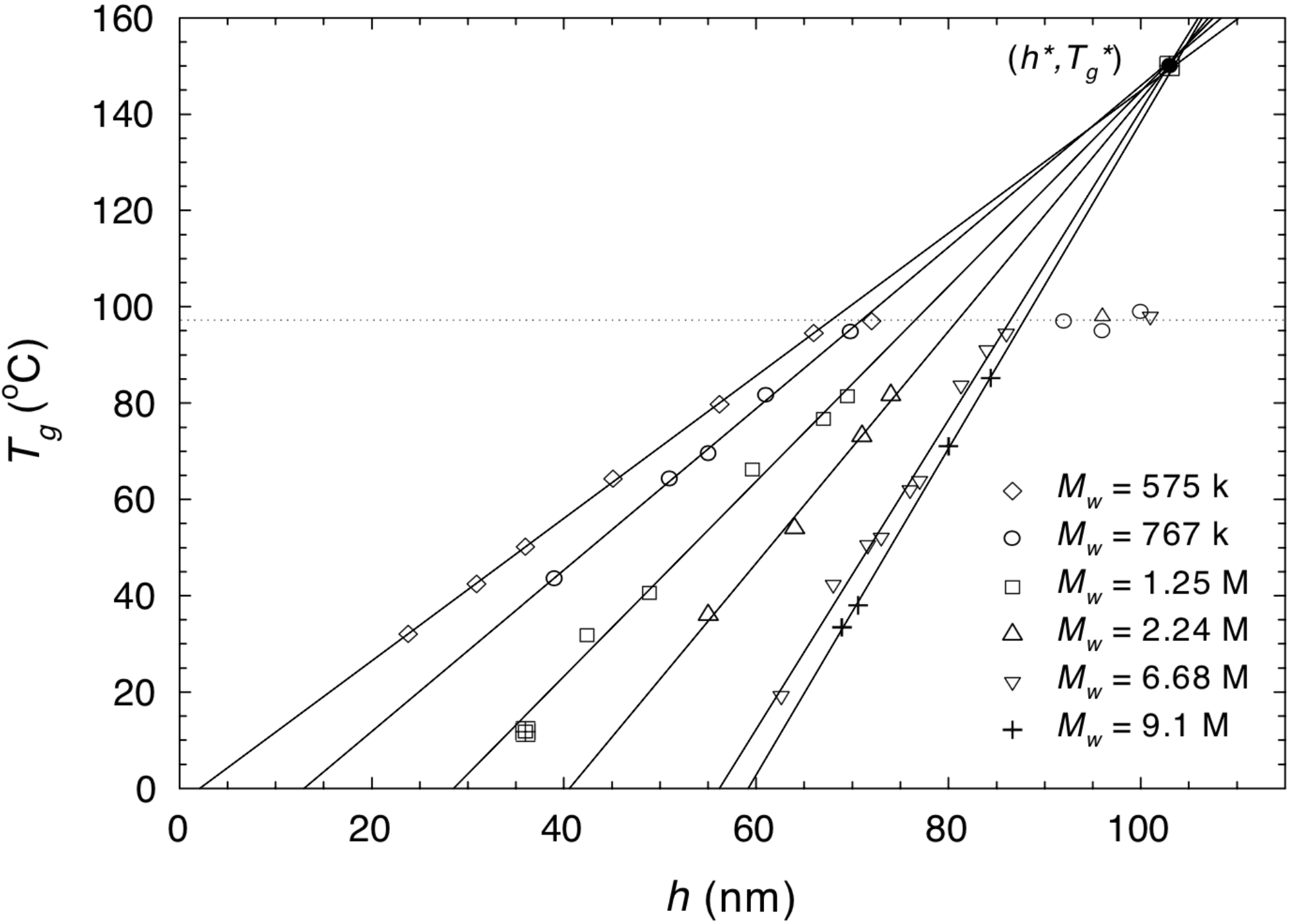}
\caption{Glass-transition temperature $T_{\textrm{g}}$ of freestanding polystyrene films as a function of film thickness $h$, for various average molecular weights $M_{\textrm{w}}$. By extrapolating the low-thickness linear regimes towards larger thicknesses, one empirically finds a universal crossing point, denoted by $(h^*,T_\text{g}^*)$. Figure reproduced from Ref.~\cite{Forrest2001}.}
\label{Fig1}
\end{figure}

Moreover, beyond the above generic confinement and interfacial behaviours of glassy materials made of small molecules or oligomers, and apart from possible residual stresses and artefacts induced by sample-preparation protocols~\cite{Reiter2005}, wether or not specific polymeric effects exist within the glassy physics is an interesting fundamental question with important practical implications given the widespread used of thin plastic films. Accordingly, freestanding polymer films with a thickness $h$ comparable to the macromolecular radius of gyration $R_{\textrm{g}}$ were experimentally studied~\cite{Dalnoki2000,Forrest2001, Roth2003,Roth2006,Kim2011}. The experiments showed that, for molecular weights $M_\text{w}<378\times10^3$ -- \textit{i.e.} $\sim 3600$ monomeric units -- $T_\text{g}$ does not exhibit any dependence on $M_\text{w}$, and the curves can be mapped onto the one for low-$M_\text{w}$ supported films. However, for larger $M_\text{w}$, the $T_\text{g}$ dependence on $h$ and $M_\text{w}$ becomes non-trivial, as shown in Fig.~\ref{Fig1}: at large thicknesses, $T_\text{g}$ is equal to the bulk value $T^{\textrm{bulk}}_\text{g}$; at small thicknesses, $T_\text{g}$ decreases in an affine way with decreasing $h$, together with a slope moderately increasing with $M_\text{w}$. As such, the glass-transition temperature follows the empirical law:
\begin{equation}
T_\text{g} = T_\text{g}^{*} +f(M_\text{w})(h-h^*),
\label{eq:empirical}
\end{equation} 
for $h<h_{\textrm{c}}$, where $h_{\textrm{c}}=h^*+(T^{\textrm{bulk}}_\text{g}-T_\text{g}^*)/f(M_\text{w})$ is a $M_\text{w}$-dependent critical thickness separating the two regimes, with $f$ a slowly increasing function of $M_\text{w}$, and where $h^*$ and $T_\text{g}^{*}$ are the coordinates of an apparent universal crossing point obtained by extrapolating the low-thickness linear regime of Eq.~\eqref{eq:empirical} towards larger thicknesses (see Fig.~\ref{Fig1}). The function $f(M_\text{w})$ was suggested to be logarithmic-like~\cite{Dalnoki2000}, a form which, perhaps coincidently, is also present in other interfacial polymeric effects~\cite{Aubouy2000}. The existence of such a sharp transition in thin supercooled polymeric films suggests a change of dominating relaxation mechanism around $h_{\textrm{c}}$, from a generic bulk molecular one above $h_{\textrm{c}}$, to a purely confinement-induced polymeric one below $h_{\textrm{c}}$ -- indicating a probable connexion between $h_{\textrm{c}}$ and some typical macromolecular polymeric length scale. In addition, the empirical trends suggest that the polymeric mechanism starts to be present below the universal onset thickness $h^*$, but remains less efficient than the bulk one for $h_{\textrm{c}}<h<h^*$. Moreover, it was proposed that the polymeric relaxation mechanism in thin supercooled polymer films requires two free interfaces to be connected by bridges consisting of individual macromolecules~\cite{Dalnoki2000} -- which we refer to as the \textit{bridge hypothesis}.

The glass transition in thin polymer films was further shown to have even a finer structure, with in fact two glass transitions occurring with some finite temperature gap in between them~\cite{Pye2011}, corroborating the existence of three competing distinct relaxation mechanisms~\cite{Tsui2008}: a bulk one, a confined molecular/monomeric one, and a confined polymeric one. These features are consistent with the observations made in Ref.~\cite{Forrest2001}, where a flow behavior in thin freestanding polymer films was only measured near the bulk value of $T_\text{g}$ -- independently of the actually-measured $T_\text{g}$. Moreover, by removing one of the two free interfaces of a freestanding polymer film, it was experimentally shown that the polymeric behaviour totally disappears~\cite{Kim2011,Baumchen2012}, which seems to corroborate the seminal bridge hypothesis. As a side remark, we note that the coupling between different relaxation mechanisms in the bulk was experimentally investigated in details recently, and revealed the role of intramolecular cooperative dynamics in the bulk polymeric glass transition~\cite{Baker2022}.
	
As an early attempt to rationalize the affine trend in Eq.~(\ref{eq:empirical}), de Gennes sketched a model based on free-volume arguments and an original \textit{sliding mechanism} involving the reptation-like propagation of stress kinks along the macromolecular bridges~\cite{PGDG2000,PGDG2000bis}. This model assumed an infinite molecular weight, as well as a Gaussian-tail distribution of the free volumes along the chain backbone, and involved an ideal-random walk scaling for the average bridge length. Despite its merits, the sliding model suffered from intrinsic limitations, and could not reproduce all the experimental observations~\cite{Kim2011}. Milner and Lipson suggested a \textit{delayed-glassification model}~\cite{Milner2010}, extending the sliding model and computing the bridge-length distribution for infinite molecular weights, that led to a depth-dependent $T_\text{g}$ and a decrease in the overall measured $T_\text{g}$~\cite{Lipson2010}. But, once again, while the qualitative picture seemed appealing, the model could not reproduce the experimental data in a quantitative fashion. 

To date, there is actually no model which quantitatively captures the $M_\text{w}$ dependence of $T_\text{g}$ in thin polymer films. In his seminal work~\cite{PGDG2000}, de Gennes suggested to refine his approach by performing a complete statistical treatment of the bridge distribution for finite-sized polymer chains in a thin film. This is thus the topic of the present article, where we compute the bridge-length distribution, its mean value, and its proportion within a film, and use the obtained results in order to critically revisit de Gennes' sliding mechanism. We note that \textit{loops}, \textit{i.e.} chain portions connecting two points of a single interface, are not considered here for two reasons. First, supported films also contain loops but do not show the $M_\text{w}$ behaviour of freestanding films~\cite{Dalnoki2000,Kim2011,Baumchen2012}. Secondly, adding loops to the calculation does not change the $M_\text{w}$ dependence.  
	
\section{Sliding mechanism}
\label{sec2}
Here, we first briefly recall the main ingredients of the sliding model~\cite{PGDG2000}. Therein, the relaxation time $\tau$ is essentially set by the time required for a \textit{kink}, \textit{i.e.} some localised stress, to travel along a \textit{bridge}, \textit{i.e.} a portion of polymer chain connecting the two free interfaces. The kink travels along the chain backbone using successive jumps involving a series of volumes $\omega_i$, which are assumed to be normally and identically Gaussian-tail distributed, as:
\begin{equation}
p(\omega_i) = \sqrt{\frac{2}{\pi\omega_0^2}}e^{-\frac{\omega_i^2}{2\omega^2_0}},
\end{equation}
where the standard deviation $\omega_0$ is assumed to be small compared to the monomer size $\sim a^3$, in order to reflect the fact that relaxation along the chain is easier than bulk molecular relaxation. The average relaxation time of this sequential process thus reads:
\begin{equation}
\tau=\tau_0 \left\langle \exp{\left(\frac{1}{v_{\textrm{f}}(T)} \sum_i \omega_i \right)}\right\rangle_{\mathcal{P}},
\end{equation}
where the average is made over the ensemble $\{\omega_i\}$ against the distributions $\mathcal{P}=\Pi_ip(\omega_i)$, and where $v_{\textrm{f}}(T)$ is the free volume at temperature $T$. As classically done, $v_{\textrm{f}}$ is assumed to vanish at a finite temperature $T_{\textrm{V}}$ and to evolve in an affine way with temperature, so that:
\begin{equation}
\label{eqfv}
v_{\textrm{f}}(T)=\alpha a^3 (T-T_{\textrm{V}}),
\end{equation}
where $\alpha$ is the expansion coefficient. Assuming that the process happens along a bridge of average number of units $\langle b \rangle$, one eventually finds:
\begin{equation}
\label{PGDG result}
T_\text{g}-T_\text{V} \propto \sqrt{\langle b \rangle}.
\end{equation}
Introducing the film thickness $h$, and assuming that $\langle b \rangle\sim h^2$ -- which is only valid for infinite ideal random walks -- then leads to the affine trend with $h$ in Eq.~(\ref{eq:empirical}). We now aim at calculating the exact bridge-length distribution for finite-sized polymer chains, in order to investigate wether or not the $M_{\textrm{w}}$ dependence in Eq.~(\ref{eq:empirical}) can also be captured by the sliding model. 
	
\section{Bridge statistics}
\label{sec3}
\begin{figure}[!h] 
\includegraphics[width=0.45\textwidth]{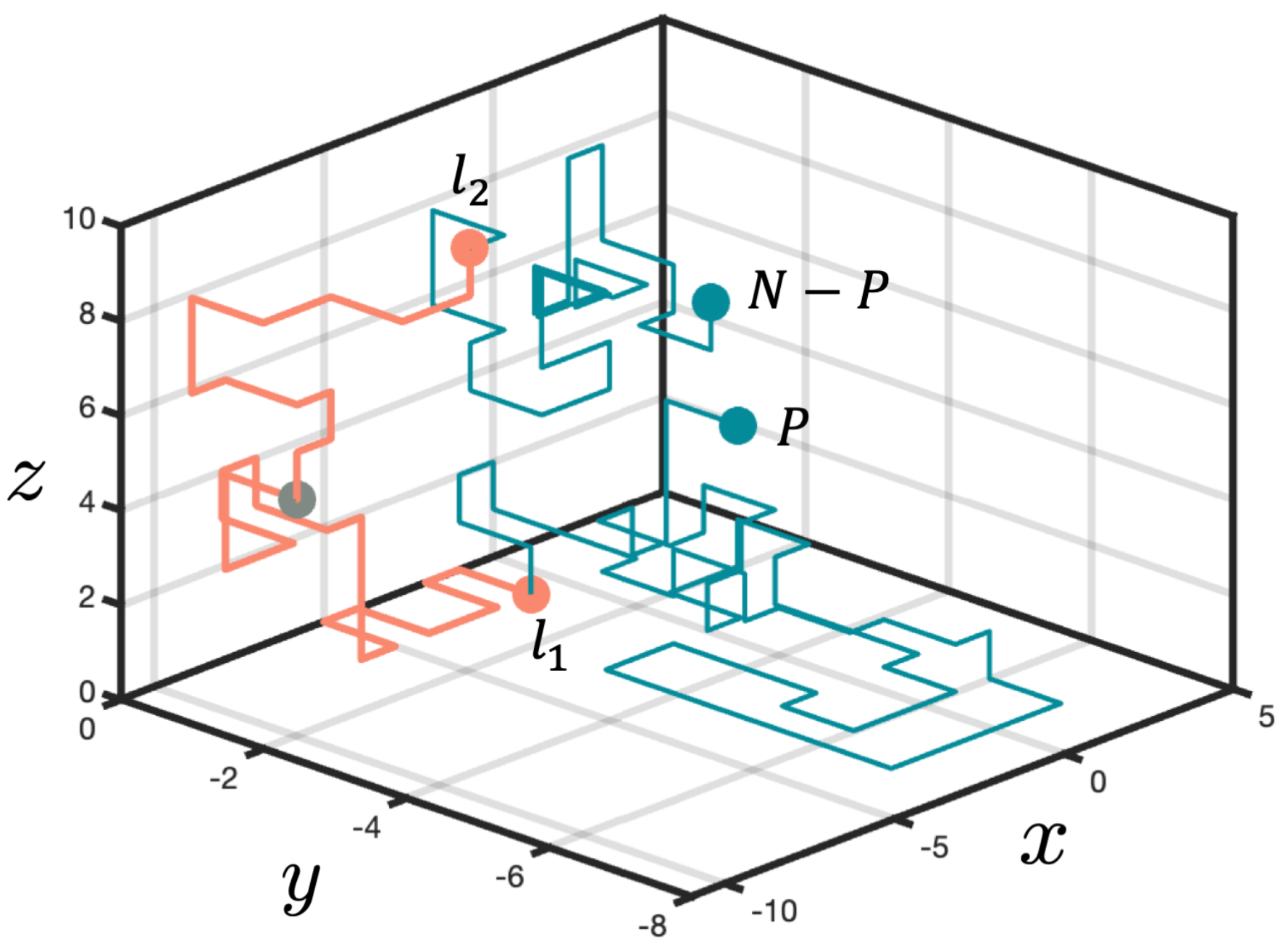}
\caption{Sketch of the problem studied, as obtained from numerical simulation. We consider an ideal random walk with $N$ steps in a box of dimensionless thickness $H=10$ along the $z$-axis. A randomly chosen monomer (grey), located at an altitude $z$, is $P$ steps away from one of the walk ends, and $N-P$ steps away from the other end. The chosen monomer is connected to both the $z=0$ and $z=H$ interfaces, with paths of dimensionless lengths $l_1$ and $l_2$, respectively, forming a bridge (orange) of dimensionless length $B=l_1+l_2$. Another part (blue) of the random walk does not belong to the bridge.}
\label{Fig2} 
\end{figure}

We consider a film made of a dense (supercooled) polymer melt consisting of identical chains, containing $N$ monomers of size $a$ (\textit{e.g.} for polystyrene, one has  a rescaled ideal monomeric size $a\approx 0.75$~nm) each. The film is assumed to be infinite in the ($x,y$)-plane, and to have two flat free interfaces located at the dimensionless vertical coordinates $z=0$ and $z=H=h/a$. We define a bridge as a segment of a polymer chain that connects the two free interfaces, as shown in Fig.~\ref{Fig2}. 

First, we are interested in the probability density of the dimensionless bridge length $B=b/a$, at a certain position $z$ inside the film. We start by picking a monomer at a distance $z$ from the lower interface. This monomer belongs to a polymer chain. From the position of the picked monomer, there are two branches of the polymer chain. As we deal with a (supercooled) polymer melt, these two branches can be properly described by Gaussian statistics~\cite{Doi1988}. For large $N$, \textit{i.e.} high molecular weight, one can invoke the continuous description of Brownian motion. As such, the probability density of the bridge length can be constructed using a constrained sum of the first-passage ``times'' of two Brownian motions. The polymer chain is of total length $N$, so that the test monomer considered above is at a distance $P$ from one end of the chain, and at a distance $N-P$ from the other end (see Fig.~\ref{Fig2}), with $P$ uniformly distributed in $\left[ 0,N\right]$. Denoting $l_1$ and $l_2$ the first-passage ``times'' of the chain from the test monomer to the $z=0$ and $z=H$ interfaces, respectively, with $l_1\leq P$ and $l_2\leq N-P$, the bridge length reads $B=l_1+l_2$.
Therefore, the probability density $\rho(N,H,B,P,z)$ that a monomer located at a distance $z$ from the bottom interface, and at position $P$ along a chain of total length $N$, belongs to a bridge of length $B$, reads:
\begin{align}
\label{eq:density_Pzb}
\rho=
2 \iint_{I_1\times I_2}\textrm{d}l_1 \textrm{d}l_2\ \delta (l_1+l_2-B)f_0(z,l_1)f_H(z,l_2),
\end{align}
with $I_1=[0,P]$, $I_2=[0,N-P]$, where the factor 2 accounts for the possibility of the two bridge subparts to be exchanged, and where $f_0(z,l_1)$ and $f_H(z,l_2)$ are the first-passage-time probability densities to the bottom and top interfaces, after ``times'' $l_1$ and $l_2$, respectively, when starting at a distance $z$ from the bottom interface~\cite{Redner2001}. By performing one of the two integrations, Eq.~(\ref{eq:density_Pzb}) becomes:
\begin{equation}
\label{eq:density_Pzb2}
\rho=2 \int_{I}\textrm{d}l f_0(z,l) f_H(z,B-l).
\end{equation}
with $I=[\max(0,B-N+P),\min(B,P)]$. 
Then, uniformly averaging Eq.~(\ref{eq:density_Pzb2}) with respect to $P$ and $z$, gives the probability density $\bar{\rho}(N,H,B)$ that a randomly chosen monomer inside the film belongs to a bridge of length $B$, between $0$ and $N$, as:
\begin{equation}
\label{eq:density_b}
\bar{\rho}=\frac{4\pi^{2}D^{2}}{H^{4}}\frac{\left(N-B\right)B}{N}\sum_{k=1}^{\infty}k^{2}\left(-1\right)^{k+1}e^{-D\lambda_{k}^{2}B},
\end{equation}
where $\lambda_k = \frac{k\pi}{H}$ and $D=\frac{1}{2d}$, with $d=3$ the space dimension. By integrating Eq.~(\ref{eq:density_b}) over $B$, from $0$ to $N$, one gets the fraction $\phi(N,H)$ of monomers belonging to bridges, as:
\begin{align}
&\phi=\frac{1}{3} - \frac{7 H^2}{90 DN}& \\
&+\frac{4}{\pi^{2}}\sum_{k=1}^{\infty} \frac{\left(-1\right)^{k+1}}{k^{2}} e^{-D\lambda_{k}^{2} N}\left( 1+\frac{2H^2}{\pi^2 D N k^2} \right),
\end{align}
where we note the expected diffusive-like self-similarity in the variable $H/\sqrt{DN}$. The limit $1/3$ at infinite $N$ is intuitive, since any sub-part of the chain containing a given monomer then touches two interfaces, either twice the top one, or twice the bottom one, or once each of the two interfaces, with equal chances.

We now turn to the central quantity of interest in this work, \textit{i.e.} the average dimensionless bridge length $\langle B\rangle$, which is a function of $N$ and $H$. It can be directly computed from Eq.~(\ref{eq:density_b}), leading to:
\begin{multline}
\label{eq:B_exact}
\langle B\rangle=\frac{1}{\phi(N,H)}\bigg[\frac{7 H^2}{90 D} -\frac{31 H^4}{1260 N D^2} \\
+\frac{4N}{\pi^{2}} \sum_{k=1}^{\infty}\frac{\left(-1\right)^{k+1}}{k^{2}}u_k(N,H) e^{-D\lambda_{k}^{2} N} \bigg],
\end{multline}
where, for convenience, we invoked the auxiliary function:
\begin{equation}
u_k =  1+\frac{4H^2}{\pi^2 D N k^2}+\frac{6H^4}{\pi^4 D^2 N^2 k^4}.
\end{equation}
By expanding Eq.~(\ref{eq:B_exact}), one finds in particular the large-$N$ asymptotic behavior:
\begin{equation}
\label{eq:B_approx}
\langle B\rangle\simeq \dfrac{7H^2}{30D}-\frac{61H^4}{3150D^2N} + O\left(\frac1{N^{2}}\right),
\end{equation}
that exhibits the $\langle B\rangle\sim H^2$ scaling invoked in the sliding model~\cite{PGDG2000}, as well as the first finite-size correction to it.

\section{Finite-size sliding mechanism}	
\begin{figure}
\centering \includegraphics[width=0.45\textwidth]{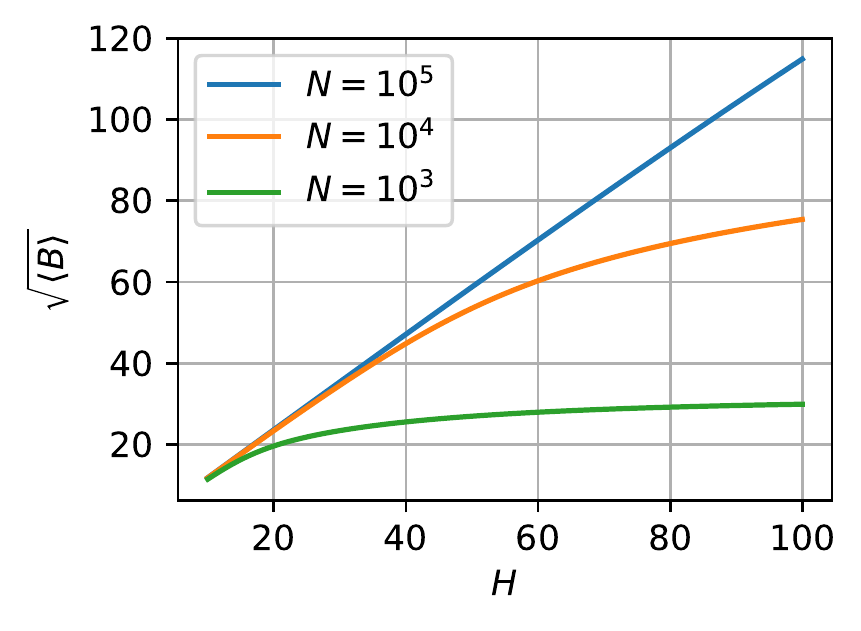}
\caption{Square root of the average dimensionless bridge length $\langle B\rangle$ as a function of dimensionless film thickness $H$, for three chain lengths $N$, obtained from the numerical evaluation of Eq.~(\ref{eq:B_exact}) using a cut-off after 100 terms in the sum. The results have been checked to weakly depend on the cut-off value in this range. }
\label{Fig3} 
\end{figure}

We now examine the modification of the sliding mechanism for polymer chains of finite length. By plugging Eq.~(\ref{eq:B_exact}) into Eq.~(\ref{PGDG result}), one can get an exact expression (not shown) for $T_\text{g}(H,N)$ from the sliding mechanism. Expanding the latter, one gets the large-$N$ asymptotic behavior: 
\begin{equation}
\label{eq:prediction}
T_\text{g}-T_\text{V} \propto \frac{H}{\sqrt{D}}\left(1-\frac{61H^2}{735DN}\right)^{1/2}.
\end{equation}
In Fig.~\ref{Fig3}, we plot $\sqrt{\langle B\rangle}$ as a function of $H$, for different values of $N$, by numerically evaluating Eq.~(\ref{eq:B_exact}). While we recover the linear behaviour introduced in Ref.~\cite{PGDG2000} in the
strong-confinement regime, the leading term in the large-$N$ asymptotics is independent of $N$. In other words -- and even without discussing the intercept -- the slope of the affine regime cannot exhibit the logarithmic-like dependence in $N$ seen in the experiments~\cite{Dalnoki2000,Forrest2001} (see Fig.~\ref{Fig1} and Eq.~\eqref{eq:empirical}), preventing the current refined sliding mechanism from explaining them, even qualitatively. We stress that Eq.~(\ref{eq:B_exact}) is essentially of the form: 
\begin{equation}
\langle B\rangle=\frac{H^2}{D}\mathcal{F}\left(\frac{H}{\sqrt{ND}}\right),
\end{equation}
with $\mathcal{F}$ a scaling function, which, combined with Eq.~(\ref{PGDG result}), cannot lead to the factorized form of Eq.~\eqref{eq:empirical}. We also stress that including loops in addition, or exclusively, does not help too, as all these types of paths can be seen under the same category of survival processes, and therefore exhibit a large-$N$ saturation of their average lengths around the value of the film thickness.
	
\section{Discussion}
\label{sec4}
In an attempt to generalize our findings to some variations in the sliding mechanism, we assume that the relaxation time is of the generalized form:
\begin{equation}
\tau = \tau_0\,\mathcal{H}\left[\langle B\rangle,\frac{a^3}{v_{\textrm{f}} (T)}\right], 
\label{eq:general}
\end{equation}
where $\tau_0$ is a reference time scale, and $\mathcal{H}$ is an increasing function of both its arguments. Indeed, the relaxation time is expected to increase with increasing bridge length or decreasing free volume. Therefore, and because the effective glass-transition temperature $T_{\textrm{g}}$ of a film is assumed to be reached when $\tau$ reaches the reference relaxation time of a bulk material at $T_{\textrm{g}}^{\textrm{bulk}}$~\cite{PGDG2000}, a relation of the following form must be satisfied:
\begin{equation}
v_{\textrm{f}} (T_{\textrm{g}}) = a^3\,\mathcal{G} \left(\langle B\rangle\right),
\label{eq:general2}
\end{equation}
where $\mathcal{G}$ is an increasing function. In the sliding model~\cite{PGDG2000}, one has $\mathcal{G}(x) \propto \sqrt{x}$ for instance. For comparison, in the delayed-glassification model~\cite{Milner2010}, one has $\mathcal{G}(x) \propto \log(x)$. As a side remark, a linear relation between free volume $v_{\textrm{f}}$ and temperature $T$ was assumed (see Eq.~\eqref{eqfv}), in view of thermal expansion in a sufficiently narrow temperature range, but we stress that any behavior of the form $v_{\textrm{f}}\sim (T-T_{\textrm{V}})^\beta$ with a positive exponent $\beta$ would lead to the same conclusion. 

Let us now exhibit a necessary condition that should be satisfied by a model to ensure its applicability for describing experimental facts. The trends in Fig.~(\ref{Fig1}) are consistent with Eq.~\eqref{eq:empirical}, and a factor $f(M_{\textrm{w}})$ slowly diverging with $M_{\textrm{w}}$~\cite{Dalnoki2000}. Assuming the latter divergence to be true implies that:
\begin{equation}
\lim_{M_{\textrm{w}}\to\infty}\frac{\partial T_{\textrm{g}}}{\partial h}=+\infty. 
\end{equation}
Combining the latter with Eq.~\eqref{eqfv}, and assuming $T_{\textrm{V}}$ and $\alpha$ to be independent of $h$, leads to:
\begin{equation} 
\lim_{N\to\infty} \frac{1}{a^3}\frac{\partial v_{\textrm{f}}(T_{\textrm{g}})}{\partial H}=+\infty.
\label{eq:criteria}
\end{equation}
Finally, by combining Eq.~\eqref{eq:criteria} with Eq~\eqref{eq:general2}, one gets:
\begin{equation} 
\lim_{N\to\infty} \mathcal{G}'(\langle B\rangle)\,\frac{\partial \langle B\rangle}{\partial H}=+\infty.
\label{eq:criteria2}
\end{equation}
From Eq.~\eqref{eq:B_approx}, we see that the left-hand side of Eq.~\eqref{eq:criteria2} equals $[7H/(15D)]\mathcal{G}'[7H^2/(30D)]$, which is positive, but finite. Hence, even a generalized formulation of the sliding model following Eq.~(\ref{eq:general}) cannot describe the experimental data. As a result, within the Gaussian framework, we can conclude that a different mechanism is needed to explain the $M_{\textrm{w}}$ dependence of the $T_\text{g}$ reductions in thin freestanding polymer films. 

\section{Conclusion}
\label{sec13}
We have computed the probability density function of the bridge length in a thin film made of a dense equilibrium assembly of identical finite-sized polymer chains. The calculations were performed in the Gaussian-chain framework. We have then used the obtained expressions in order to refine and critically discuss the sliding model for the anomalous glass transition in thin freestanding polymer films. Our analysis suggests that the sliding model, as well as similar models based on free-volume arguments, can not capture the intricate chain-length dependence of the experimental data. Another key physical ingredient, with a dependence on the molecular weight, seems to be missing. Finally, we note that: i) the remarkable stability of the films above the measured $M_\text{w}$-dependent $T_\text{g}$; ii) the proposed existence of a second $T_\text{g}$, more closely associated with flow; and iii) the coincidence of $T_\text{g}^*$ with the temperature of the $\alpha-\beta$ splitting in polystyrene, may all suggest that the $M_\text{w}$-dependent $T_\text{g}$ in freestanding polymer films is associated with a local, rather than segmental, relaxation. 
\newline

\section*{Acknowledgments}
The authors acknowledge financial support from the European Union through the European Research Council under EMetBrown (ERCCoG-101039103) grant. The authors also acknowledge financial support from the Agence Nationale de la Recherche under EMetBrown (ANR-21-ERCC-0010-01), Softer (ANR-21-CE06-0029) and Fricolas (ANR-21-CE06-0039) grants, as well as from the UHJ-France association and the Scopus Foundation. Finally, they thank the Soft Matter Collaborative Research Unit, Frontier Research Center for Advanced Material and Life Science, Faculty of Advanced Life Science at Hokkaido University, Sapporo, Japan, as well as the Natural Sciences and Engineering Research Council of Canada. 
	
\section*{Author contribution statement}
J.F., E.R. and T.S. conceived the study. H.B. and M.A. performed the analytical research. R.C. performed the numerical simulations. H.B. wrote the first draft of the manuscript. All the authors discussed the results and contributed to the writing of the manuscript.

\section*{Data availability statement}
Data produced for this article are available upon reasonable request to the authors.

\end{document}